\documentclass[10pt,conference]{IEEEtran}
\usepackage{amssymb,latexsym,pb-diagram}

\newcommand{\QQ}{\mathbb{Q}}

\newcommand{\Ac}{\mathcal{A}}
\newcommand{\Cc}{\mathcal{C}}

\newcommand{\xm}{\mathbf{X}}

\newcommand{\im}{\mathbf{I}}

\newcommand{\Mc}{\mathcal{M}}
\newcommand{\Gal}{\mbox{Gal}}

\newtheorem{definition}{Definition}
\newtheorem{prop}{Proposition}
\newtheorem{example}{Example}
\newtheorem{remark}{Remark}
\newtheorem{lemma}{Lemma}

\newtheorem{corollary}{Corollary}

\begin{document}

%*************************** paper title***********************************%

\title{Families of unitary matrices \\ achieving full diversity}

%************** author names and affiliations******************************%

\author{\authorblockN{Fr\'ed\'erique Oggier}
\authorblockA{
Ecole Polytechnique F\'ed\'erale de Lausanne\\
Laboratoire de math\'ematique algorithmique \\
1015-Lausanne, Switzerland \\
Email: frederique.oggier@epfl.ch}
\and
\authorblockN{Emmanuel Lequeu}
\authorblockA{Mathematisches Institut  \\
Georg-August Universit\"at  \\
37073 G\"ottingen, Germany\\
Email:lequeu@uni-math.gwdg.de }
}
\maketitle

\begin{abstract}
This paper presents an algebraic construction of families of unitary matrices 
that achieve full diversity. They are obtained as subsets of cyclic division 
algebras.
\end{abstract}

%**************************************************************************%
%
% PROBLEM STATEMENT
%
%**************************************************************************%

\section{Problem statement}\label{sec:prob_stat}

In the context of noncoherent multiple antennas channel coding, 
research has been done on constructing families of unitary matrices 
with {\em full diversity}, that is, satisfying that the determinant of 
the difference of any two matrices in the family is nonzero. Among the 
algebraic approaches to this problem, the theory of fixed-point-free 
groups and their representations has been exploited in \cite{SHHS},    
while representations of Lie groups has been investigated 
by Jing and Hassibi (see e.g. \cite{Jing}). 

At the same time, division algebras for space-time coding have been 
introduced in the context of coherent MIMO systems \cite{Sethuraman}. 
These algebras became of great interest, since they naturally provide a 
linear family of fully-diverse matrices.

The aim of this work is to show that division algebras (in particular cyclic 
division algebras) can also be used 
to construct fully diverse unitary matrices. 

The paper is organized as follows. In the next section, we recall 
the basic facts about cyclic algebras. In section \ref{sec:unitary}, 
we explain how the condition of being unitary for a matrix can be 
translated into first a constraint on an element of the algebra, and second 
a constraint on a commutative subfield of the algebra. 
This contains a constructive proof that yields a way of exhibiting unitary 
matrices. The whole process is illustrated in a worked out example in 
section \ref{sec:fdum}.

\begin{remark}\rm
In the following, we choose the dimension of the algebra to be 3 
for the sake of simplicity. The same theory can be generalized 
for any dimension $n$.  
\end{remark}

%**************************************************************************%
%
% CUBIC CYCLIC ALGEBRAS
%
%**************************************************************************%

\section{Cubic cyclic algebras}\label{sec:cca}

In this section, we briefly recall what is a cubic cyclic algebra, and 
how it provides a linear family of $3 \times 3$ fully-diverse matrices. 
Let $L$, $K$ be two number fields.

%**************************************************************************%
%
\subsection{The algebra structure}

Let $L/K$ be a Galois extension of degree $3$ such that its Galois 
group $G=\Gal(L/K)$ is cyclic, with generator $\sigma$. 
Namely, $G=\{\sigma,\sigma^2,\sigma^3= Id \}$. 
Such an extension is called {\em cyclic}.
Denote by $K^*$ (resp. $L^*$) the set of non-zero elements of $K$ (resp. $L$). 
We choose an element $\gamma \in K^*$. We construct a non-commutative 
algebra, denoted $\Ac=(L/K,\sigma,\gamma)$, as follows:
\[
\Ac= L \oplus e L \oplus e^2 L 
\]
such that $e$ satisfies
\[
e^3 = \gamma ~~\mbox{ and }~~ \lambda e = e \sigma(\lambda)  
\mbox{ for } \lambda \in L.
\]
Such an algebra is called {\em a cubic cyclic algebra}. It is a right vector 
space over $L$, and as such has dimension $(\Ac : L)= 3$.

Cubic cyclic algebras naturally provide linear families of matrices thanks to 
an isomorphism between the {\em split} algebra $\Ac \otimes_K L$ 
and the algebra $\Mc_3(L)$, the $3$-dimensional matrices with coefficients 
in $L$. This isomorphism, denote it by $h$, is given explicitly.
Since each $x \in \Ac$ is expressible as
\[
x=x_0+e x_1+e^2 x_2,~x_i \in L \mbox{ for all }i,
\]
it is enough to give $h(x_i \otimes 1)$ and $h(e \otimes 1)$.
We have that 
\begin{equation}\label{eq:h}
h:\Ac \otimes_K L \cong \Mc_3(L)
\end{equation}
is given by, for all $i$, 
\[
x_i \otimes 1 \mapsto 
\left( \begin{array}{ccc}
        x_i & 0             & 0  \\
         0  & \sigma(x_i)   & 0  \\
         0  &    0          & \sigma^2(x_i) \\
       \end{array} \right),~ 
e \otimes 1 \mapsto 
\left( \begin{array}{ccc}
         0      &  0  &   \gamma \\
         1      &  0  &   0      \\
         0      &  1  &   0    \\
        \end{array} \right). 
\]
Thus the matrix of $h(x \otimes 1)$ is easily checked to be 
\begin{equation}
\left(\begin{array}{ccc}
x_0 & \gamma\sigma(x_2) & \gamma\sigma^2(x_1)  \\
x_1 & \sigma(x_0)       & \gamma\sigma^2(x_2)  \\
x_2 & \sigma(x_1)       & \sigma^2(x_0)         \\
\end{array}\right).
\label{eq:matrix}
\end{equation}
\begin{remark}\rm \label{rem:mult_alpha}
Notice that (\ref{eq:matrix}) is also the matrix of left multiplication by 
$x$ in the basis $\{1,e,e^2\}$.
\end{remark}
We thus start with the family of matrices
\[
\Cc=\left\{ \xm= \left(\begin{array}{ccc}
x_0 & \gamma\sigma(x_2) & \gamma\sigma^2(x_1)  \\
x_1 & \sigma(x_0)       & \gamma\sigma^2(x_2)  \\
x_2 & \sigma(x_1)       & \sigma^2(x_0)         \\
\end{array}\right)| x_0,x_1,x_2 \in L \right\},
\] 
which is linear, since it has an algebra structure.

%**************************************************************************%
%
\subsection{The diversity property}

Recall that the diversity product $\zeta(\Cc)$ of a set $\Cc$ of 
$M$ unitary $3 \times 3$ matrices $\xm_1,\ldots,\xm_M$ is the minimal 
diversity distance 
\[
\zeta(\Cc):=\frac{1}{2}\min_{i \neq j}|\det(\xm_i-\xm_j)|^{1/3}.
\]
A set of matrices with $\zeta(\Cc)>0$ is said to have {\em full diversity}. 
If $\Cc$ is linear, then the above definition simplifies to
\[
\zeta(\Cc):=\frac{1}{2}\min_{\xm \in \Cc \neq {\bf 0}}|\det(\xm)|^{1/3},
\]
in which case full diversity is obtained if all matrices are invertible.
Take now $\Cc \subset \Ac$, $\Ac$ an algebra.
If furthermore $\Ac$ is a {\em division algebra}, all matrices in $\Cc$  
are by definition invertible. Thus $\Cc$ is fully diverse.
Summarizing, if there is $\Cc$ a subset of unitary matrices in $\Ac$ 
a cyclic division algebra, then this family will automatically be fully 
diverse. 

To decide whether a cyclic algebra is a division algebra, the 
following criterion is available:
\begin{prop}\cite[p.~279]{Pierce}\label{prop:cycl_division}
Let $L/K$ be a cyclic extension of degree $n$ with Galois group 
$\Gal(L/K)=<\sigma>$. If $\gamma$ and its powers 
$\gamma^2,\ldots,\gamma^{n-1}$ are not a norm, then $(L/K,\sigma,\gamma)$ 
is a division algebra. 
\end{prop}

%**************************************************************************%
%
% THE UNITARY CONSTRAINT
%
%**************************************************************************%

\section{The unitary constraint}\label{sec:unitary}

Suppose $\Ac$ is a division algebra. We now have a linear family of 
invertible matrices among which we are looking for unitary matrices, i.e., 
$\xm \in \Cc$ such $\xm \xm^*= \im_3$, where $^*$ denotes the transpose 
conjugate.
%
%*************************************************************************%
%
\subsection{The unitary constraint in the algebra}

\begin{figure}
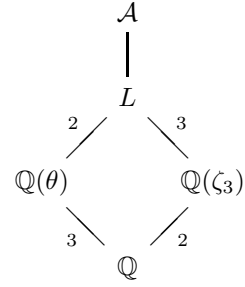

\[
\divide\dgARROWLENGTH by2
\begin{diagram}
\node[2]{\Ac}\arrow{s,-}\\
\node[2]{L} \arrow{sw,l,-}{2}\arrow{se,l,-}{3} \\
\node{\QQ(\theta) }\arrow{se,r,-}{3}\node[2]{\QQ(\zeta_3)}\arrow{sw,r,-}{2}\\
\node[2]{\QQ}\\
\end{diagram}
\]
\caption{The cyclic algebra $\Ac=(L/K,\sigma,\gamma)$.}
\label{fig:compositum}
\end{figure}

We take advantage of the matrices coming from the algebra $\Ac$ and  
use the following correspondances:
\[
x \in \Ac \hookrightarrow x \otimes 1 \in \Ac \otimes_K L 
\mapsto \xm \in \Mc_3(L)
\]
We thus translate the condition of ``being unitary'' for the matrix 
$\xm$ into a condition on the element $x$ in the algebra. 
We will show that $\Ac$ can be endowed with an {\em involution} 
$\alpha$, and that 
\[
\xm \xm^*= \im_3 \iff h(x \otimes 1)h(x \otimes 1)^*=1 \iff  x \alpha(x)=1.
\]
\begin{remark}
Defining an involution on the algebra (i.e., a map that satisfies the three 
properties described in the proof of Proposition \ref{prop:alphaL}) 
and checking that it is well defined 
on the algebra of matrices is technical, but this formalism is required 
to be sure that our objects are well defined. This subsection ends with 
an example that illustrates the theory.
\end{remark}
Let us first define an involution on $\Ac$.
\begin{prop}\label{prop:alphaL}
Let $\alpha_L: L \rightarrow L$ be an involution on $L$ such that 
$\alpha_L$ commutes with all elements of $\Gal(L/K)$.
Let $z=\gamma\alpha_L(\gamma)$ and 
$\alpha: \Ac \rightarrow \Ac$ such that 
\begin{eqnarray*}
\alpha(x_0+e x_1+e^2 x_2) &= &\alpha_L(x_0)+e^{-1}z\sigma^{-1}(\alpha_L(x_1))\\
                          & & + e^{-2}z^2\sigma^{-2}(\alpha_L(x_2)).
\end{eqnarray*}
Then $\alpha$ defines an involution on $\Ac$.
\end{prop}
\begin{remark}\label{rem:stabk}
Note that the condition that $\alpha_L$ commutes with all elements 
of $\Gal(L/K)$ implies that $\alpha_L(K)=K$. Indeed,
\[
\sigma(\alpha_L(k))=\alpha_L(\sigma(k))=\alpha_L(k), 
\mbox{ for any } k \in K,
\]
showing that $\alpha_L(k)$ is fixed by $\sigma$. In particular, 
$z \in K$.
\end{remark}
\begin{proof}
Note first that we have $e \alpha(e)=z$.
Check that
\begin{enumerate}
\item 
$\alpha(x+y)=\alpha(x)+\alpha(y)$ for all $x,y \in \Ac$.\\ 
This is clear.
\item 
$\alpha(e^jy_j e^i x_i)=\alpha(e^ix_i)\alpha(e^jy_j)$ for $x_i,y_j \in L$.\\
If $i+j < 3$, we have
\begin{eqnarray*}
\alpha(e^jy_j e^i x_i) & = & \alpha(e^{i+j}\sigma^i(y_j)x_i) \\
                   & = & e^{-(i+j)}z^{i+j}\sigma^{-(i+j)}
                        (\alpha_L(x_i)\alpha_L(\sigma^i(y_j))) \\
                   & = & e^{-(i+j)}z^{i+j}\sigma^{-(i+j)}
                        (\alpha_L(x_i))\sigma^{-j}(\alpha_L(y_j)). 
\end{eqnarray*}
Now, the right handside term is given by
\begin{eqnarray*}
\alpha(e^ix_i)\alpha(e^jy_j)&\!\!\!=\!\!\!&e^{-i}z^i\sigma^{-i}(\alpha_L(x_i)) 
                                          e^{-j}z^j\sigma^{-j}(\alpha_L(y_j))\\
                         &\!\!\!=\!\!\! & e^{-(i+j)}z^{i+j}\sigma^{-(i+j)}
                                     (\alpha_L(x_i))\sigma^{-j}(\alpha_L(y_j)).
 \end{eqnarray*}
Similarly, if $i+j \geq 3$, $i+j =3+k$, $0 \leq k \leq 2$ and the same 
computations hold.
\item $\alpha(\alpha(x))=x$ for all $x \in \Ac$.\\
      We have
      \begin{eqnarray*}
      \alpha(\alpha(e^ix_i))& =& \alpha(e^{-i}z^i\sigma^{-i}(\alpha_L(x_i)))\\ 
                           & = & \alpha_L(\sigma^{-i}(\alpha_L(x_i))
                                 \alpha_L(z^i) \alpha(e^{-i})\\
                           & = & \sigma^{-i}(x_i)\alpha_L(z^i)z^{-i}e^i. \\ 
      \end{eqnarray*}
Since $z$ is fixed by $\alpha_L$, we get that
\[
\alpha(\alpha(e^ix_i))= e^ix_i.
\] 
\end{enumerate}
\end{proof}
The involution $\alpha$ defined in the above proposition is extended to 
the split algebra $\Ac \otimes_K L \cong \Mc_3(L)$ as follows.
\[
\alpha \otimes \alpha_L: \Ac \otimes_K L  \rightarrow \Ac \otimes_K L.
\]
It is used to define an involution $\alpha_h$ on $\Mc_3(L)$ via the 
isomorphism $h$:
\begin{equation}\label{eq:alphah}
\alpha_h = h \circ (\alpha \otimes \alpha_L) \circ  h^{-1}.
\end{equation}
\begin{prop}
Let $\xm=h(x \otimes 1)$ and $z$ be as in the hypothesis of the previous 
proposition. 
If $z=1$, then $\alpha_h(\xm)=\xm^*$.
\end{prop}
\begin{proof}
Recall first that 
\[
\xm=\left(\begin{array}{ccc}
x_0 & \gamma\sigma(x_2) & \gamma\sigma^2(x_1)  \\
x_1 & \sigma(x_0)       & \gamma\sigma^2(x_2)  \\
x_2 & \sigma(x_1)       & \sigma^2(x_0)         \\
\end{array}\right).
\]
We have
\begin{eqnarray*}
\alpha_h(\xm) & = & \alpha_h(h(x \otimes 1 )) \\
              & = & h \circ (\alpha \otimes \alpha_L)(x \otimes 1 )\\
              & = & h(\alpha(x) \otimes \alpha_L(1))\\
              & = & h(\alpha(x) \otimes 1) \alpha_L(1).
\end{eqnarray*}
Recall that $e^{-1}=\gamma^{-1}e^2$, $\gamma^{-1}=z^{-1}\alpha_L(\gamma)$ and 
that $h(\alpha(x) \otimes 1)$ is the matrix of multiplication by 
$\alpha(x)$ (see Remark \ref{rem:mult_alpha}).
Since
\begin{eqnarray*}
\alpha(x) & = & \alpha_L(x_0) +e^{-1}z\sigma^{-1}(\alpha_L(x_1))
                +e^{-2}z^2\sigma^{-2}(\alpha_L(x_2)) \\
          & = & \alpha_L(x_0) + e\gamma^{-1}z^2\sigma(\alpha_L(x_2)) 
                + e^2\gamma^{-1}z\sigma^2(\alpha_L(x_1)),
\end{eqnarray*}
we have
\[
\begin{array}{ll}
h(\alpha(x) \otimes 1) = &  \\
\!\!\!\!\left(\begin{array}{ccc}
\alpha_L(x_0) & z\alpha_L(x_1)   & z^2 \alpha_L(x_2)  \\
\alpha_L(\gamma)z\sigma(\alpha_L(x_2)) & \sigma(\alpha_L(x_0))&z\sigma(\alpha_L(x_1))   \\
\alpha_L(\gamma)\sigma^2(\alpha_L(x_1)) & \alpha_L(\gamma)z\sigma^2(\alpha_L(x_2)) & \sigma^2(\alpha_L(x_0)) \\
\end{array}\right)\!\!. & 
\end{array}
\]
Since $z=1$, $\alpha_L$ and $\sigma$ commute, and $\alpha_L$ is 
multiplicative, we get the desired result.
\end{proof}
Notice that clearly $z=1$ is the only possible choice for the involution 
$\alpha_h$ to be the conjugate transpose.
These formal proofs finally yield the desired result:
\begin{corollary}\label{cor:equiv}
We have the following equivalence:
\[
\xm \xm^*= \im_3 
%\iff (x \otimes 1)(\alpha_L \otimes \alpha)(x \otimes 1)^*= WHAT 
\iff  x \alpha(x)=1.
\]
\end{corollary}
%
%\begin{proof}
%Since 
%\begin{eqnarray*}
%\xm \xm^* & = & h(x \otimes 1)\alpha_h(h(x \otimes 1)) \\
%          & = & h(x \otimes 1)h \circ (\alpha_L \otimes \alpha)(x \otimes 1)),
%          \mbox{ using }(\ref{eq:alphah}),
%\end{eqnarray*} 
%the first equivalence comes from $h$ being an isomorphism.
%\end{proof}
%
%
\begin{example}\rm \label{ex:alpha}
Let $K=\QQ(\zeta_3)$ be a cyclotomic field, where $\zeta_3$ is a primitive 
third root of unity, and let $L=K\QQ(\theta)$ be the compositum of $K$ and 
a totally real cubic number field $\QQ(\theta)$, with discriminant 
coprime to the discriminant of $K$ and cyclic Galois 
group $\Gal(\QQ(\theta)/\QQ)=<\sigma>$ 
(see Figure \ref{fig:compositum}). 
We consider the algebra $\Ac=(L/K,\sigma,\gamma)$, where $\gamma= \zeta_3$.

The involution on $L$ is given by
\[
\begin{array}{llcl}
\alpha_L: & L & \rightarrow & L \\
          &  a_0 + a_1\theta + a_2\theta^2 & \mapsto & 
            \tau(a_0)+\tau(a_1)\theta+\tau(a_2)\theta^2 
\end{array}
\]
where $\tau$ is the generator of the Galois group $\Gal(\QQ(\zeta_3)/\QQ)$. 
Namely, $\tau(b_0+b_1\zeta_3)=b_0+b_1\zeta_3^2$ and $\zeta_3^2=-\zeta_3-1$.
The involution $\alpha_L$ satisfies that $\tau$ commutes with $\sigma$,
so that the involution 
\[
\begin{array}{llcl}
\alpha: & \Ac & \rightarrow & \Ac   \\ 
        & x=x_0+e x_1+e^2 x_2 & \mapsto & \alpha(x) \\
\end{array}
\]
where $\alpha(x)= \alpha_L(x_0)+e \zeta_3^2 \sigma(\alpha_L(x_2))
+e^2 \zeta_3^2\sigma^2(\alpha_L(x_1))$
is well defined by Proposition \ref{prop:alphaL}. \\
We have
\[
\begin{array}{c}
\left(\begin{array}{ccc}
0 & 0 & \gamma \\
1 & 0 & 0 \\
0 & 1 & 0  \\
\end{array}\right)
\left(\begin{array}{ccc}
0            & 1 & 0 \\
0            & 0 & 1 \\
\bar{\gamma} & 0 & 0  \\
\end{array}\right)=
\left(\begin{array}{ccc}
1 & 0 & 0 \\
0 & 1 & 0 \\
0 & 0 & 1  \\
\end{array}\right) \\
\iff  \gamma\bar{\gamma}=1 \iff e \alpha(e)=1.
\end{array}
\]
\end{example}
%
%
%*************************************************************************%
%
\subsection{The unitary constraint in commutative subfields}

We now show how the problem of finding unitary elements in the algebra 
$\Ac$ can be reduced to find elements of norm 1 in commutative subfields 
of $\Ac$.
\begin{prop}
Let $\Ac=(L/K,\sigma,\gamma)$ be a cyclic division algebra, 
and $ x \in \Ac$ such that $x \neq \pm 1$.
The following statements are equivalent:
\begin{enumerate}
\item $x \alpha(x)=1$.
\item There exists $u \in \Ac^*$ such that $u \alpha(u)=\alpha(u) u$ 
      and $x=u \alpha(u)^{-1}=\alpha(u)^{-1} u$.
\end{enumerate}
\end{prop}
\begin{proof}
The sufficient condition is clear. Let us prove the necessary condition. 
Assume that $x \alpha(x)=1$. Let $M$ denote the 
subfield of $\Ac$ generated by $K$ and $x$. It is commutative and 
satisfies that $\alpha(M)=M$, since $\alpha(K)=K$ and $\alpha(x)=x^{-1}$.
Thus
\[
M^{\alpha}=\{ y \in M ~|~ \alpha(y)=y \},
\]
the subfield of $M$ fixed by $\alpha$ is well-defined.\\ 
{\em Claim:} $M/M^{\alpha}$ is a quadratic extension with Galois group 
$\Gal(M/M^{\alpha})=\{ Id_M, \alpha|_M \}$. \\
It is enough to prove that $\alpha|_M \neq Id_M$. If $\alpha|_M=Id_M$, 
then $x\alpha(x)=x^2=1$, implying that $x = \pm 1$, which is a contradiction.

The condition $x\alpha(x)=1$ becomes $N_{M/M^{\alpha}}(x)=1$. 
By a corollary of 
Hilbert 90 Theorem, there exists $u \in M^*$ such that 
$x = u/\alpha(u)$.
\end{proof}
The above proof gives a way of building unitary elements of 
the algebra $\Ac$. Take a commutative subfield $M$ of $\Ac$ such that 
$\alpha(M)=M$ but with $y \in M$ such that $\alpha(y) \neq y$, 
so that $M^{\alpha}$ is not $M$ itself. Take $u \in M^*$ and compute 
$x = u/\alpha(u)$. The element $x \in \Ac$ will satisfy 
$x \alpha(x)=1$.

The next step is thus how to build commutative subfields of $\Ac$ which 
are stable by $\alpha$. 
\begin{definition}\label{def:red_char_pol}\cite{Reiner}[p.~113]
Let $\Ac$ be a cubic cyclic algebra. For $x \in \Ac$, define its 
{\em reduced characteristic polynomial} $\chi_x$ as the characteristic 
polynomial of $h(x \otimes 1)$.
\end{definition}
Let $x \in \Ac$. Its reduced characteristic polynomial is given by 
\begin{eqnarray*}
\chi_x(X) & = & \det
              \left(\begin{array}{ccc}
              x_0-X & \gamma\sigma(x_2)   & \gamma\sigma^2(x_1)  \\
              x_1   & \sigma(x_0)-X       & \gamma\sigma^2(x_2)  \\
              x_2   & \sigma(x_1)         & \sigma^2(x_0)-X         \\
              \end{array}\right) 
%        & = & -X^3 + ...
\end{eqnarray*}
It is shown that for each $x \in \Ac$, its reduced characteristic 
polynomial lies in $K[X]$ \cite{Reiner}[p.~113]. 
\begin{lemma}\label{lemma:irr}
If $x \not\in K$, the center of $\Ac$, then $\chi_x$ is irreducible over $K$.
\end{lemma}
\begin{proof}
Either $\chi_x$ splits into three linear terms in $K[X]$, which contradicts 
the hypothesis that $x \not\in K$, or it factors into one linear term 
in $K[X]$ and one quadratic term irreducible over $K$. But $(\Ac:K)=9$, 
so that $\Ac$ cannot contain a quadratic subfield.
\end{proof}
If $x \not \in K$, $K[X]/(\chi_x(X))$ is a 
commutative subfield of $\Ac$, of degree 3 over $K$.
\begin{example}\rm \label{ex:K(e)}
Let $\Ac=(L/K,\sigma,\gamma)$ and $\Ac=L \oplus eL \oplus e^2L$.
Let $\chi_e$ be the reduced characteristic polynomial of $e$, given by
\begin{eqnarray*}
\chi_e(X) & = & \det
              \left(\begin{array}{ccc}
              -X &  0   & \gamma  \\
               1 & -X   & 0  \\
               0 &  1   & -X         \\
              \end{array}\right) \\
        & = & -X^3+\gamma.
\end{eqnarray*}
By Lemma \ref{lemma:irr}, $\chi_e$ is irreducible and 
$K[X]/\chi_e(X) \cong K(e)$ is a commutative subfield of $\Ac$.
\end{example}
\begin{remark}\rm
We have not discussed whether the commutative subfields that we build 
contain a quadratic subfield fixed by $\alpha$. This will be illustrated  
later in the example of Section \ref{sec:fdum}. 
\end{remark}

%**************************************************************************%
%
% FAMILIES OF FULLY-DIVERSE UNITARY MATRICES 
%
%**************************************************************************%

\section{Families of fully-diverse unitary matrices}\label{sec:fdum}

In this section, we consider a particular cyclic division algebra 
and show how to use it to build families of fully-diverse unitary 
matrices. 

Let $K=\QQ(\zeta_3)$ and $L=\QQ(\zeta_7+\zeta_7^{-1},\zeta_3)$ be the 
compositum of $\QQ(\zeta_3)$ and $\QQ(\zeta_7+\zeta_7^{-1})$, the 
maximal real subfield of the cyclotomic field $\QQ(\zeta_7)$ 
(see Figure \ref{fig:compositum}, with $\theta=\zeta_7+\zeta_7^{-1}$).\\
We have $\Gal(L/\QQ(\zeta_3))=
\langle \sigma \rangle$, with 
$\sigma:\zeta_7+\zeta_7^{-1} \mapsto \zeta_7^2+\zeta_7^{-2}$.

Let $\Ac =(\QQ(\zeta_7+\zeta_7^{-1},\zeta_3)/\QQ(\zeta_3),\sigma,\zeta_3)$ 
be the corresponding cyclic algebra.
This is a division algebra \cite{perfect_codes}.
As already explained in Example \ref{ex:alpha}, the involution $\alpha$ 
on $\Ac$ is given by
\[
\begin{array}{llcl}
\alpha: & \Ac & \rightarrow & \Ac   \\ 
        & x=x_0+e x_1+e^2 x_2 & \mapsto & \alpha(x),
\end{array}
\]
where $\alpha(x)=\alpha_L(x_0)+e \zeta_3^2 \sigma(\alpha_L(x_2))
+e^2 \zeta_3^2\sigma^2(\alpha_L(x_1))$ and 
$\alpha_L(b_0+b_1\zeta_3)=b_0+b_1\zeta_3^2$.\\
Note that $\alpha_L$ is here given by the usual complex conjugation.
%
%****************************************************************************%
%
\subsection{Commutative subfields of $\Ac$}

\begin{figure}
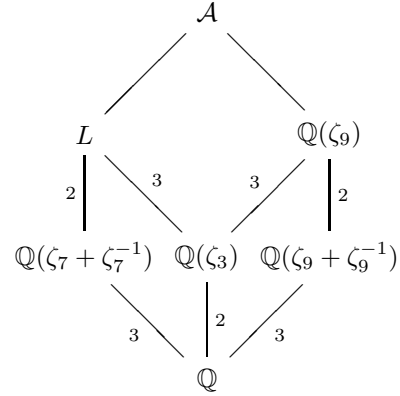

\[
\divide\dgARROWLENGTH by2
\begin{diagram}
\node[2]{\Ac} \arrow{sw,-}\arrow{se,-} \\
\node{L}  \arrow{s,l,-}{2}\arrow{se,l,-}{3}
\node[2]{\QQ(\zeta_9)} \arrow{sw,l,-}{3}\arrow{s,r,-}{2} \\
\node{\QQ(\zeta_7+\zeta_7^{-1} )} \arrow{se,r,-}{3}
\node{\QQ(\zeta_3 )}\arrow{s,r,-}{2}
\node{\QQ(\zeta_9+\zeta_9^{-1} )} \arrow{sw,r,-}{3} \\
\node[2]{\QQ}
\end{diagram}
\]
\caption{The algebra $\Ac$ and some of its commutative subfields}
\label{fig:subfields}
\end{figure}

We consider the construction of commutative subfields of $\Ac$ which
are stable by $\alpha$. 
The first obvious subfield of $\Ac$ one can think of is $L$. By definition, 
it contains a totally real quadratic subfield, 
namely $\QQ(\zeta_7+\zeta_7^{-1})$. 
Then, as explained in Example \ref{ex:K(e)}, we can consider $K(e)$, 
with minimal polynomial $\chi_e(X)=X^3-\zeta_3$. Thus $K(e)=\QQ(\zeta_9)$, 
with maximal real subfield $\QQ(\zeta_9+\zeta_9^{-1})$
(See Figure \ref{fig:subfields}).

Let us now try to determine more systematically which are the commutative 
subfields $M$ of $\Ac$ such that $M^{\alpha}\neq M$.
Clearly $M^{\alpha} \subseteq \{x \in \Ac ~|~ \alpha(x)=x \}$. 
We thus look for conditions so as to satisfy $x=\alpha(x)$.
\begin{lemma}\label{lemma:cond_coeff}
Let $x=x_0+e x_1 + e^2 x_2$, with $x_i \in L$, that is 
$x_i = v_i + \zeta_3 w_i$, $v_i$, $w_i \in \QQ(\zeta_7+\zeta_7^{-1})$ 
for $i=0,1,2$.
We have
\begin{eqnarray*}
x = \alpha(x) & \iff & \left\{ \begin{array}{ccl}
                                  x_0 & = & \alpha_L(x_0) \\
                                  v_1 & = & - \sigma(v_2) \\
                                  w_1 & = & \sigma(w_2)+v_1
                               \end{array} \right.
\end{eqnarray*}
\end{lemma}
\begin{proof}
This is a straightforward computation.
Identify the coefficients of the power of $e$
\[
\left\{ \begin{array}{ccl}
x_0 & = & \alpha_L(x_0) \\
x_1 & = & \alpha_L(\sigma(x_2))\gamma^{-1}  \\
x_2 & = & \alpha_L(\sigma^2(x-1))\gamma^{-1}
\end{array} \right.
\]
then develop and using that $\gamma = \zeta_3$, identify the constant term 
and the coefficient of $\zeta_3$.
\end{proof}
\begin{example}\rm
Let $x = x_0 +e x_1 + e^2 x_2 \in \Ac$, with $x_i = v_i + \zeta_3 w_i$, 
$v_i$, $w_i \in \QQ$ for $i=0,1,2$. The conditions of Lemma 
\ref{lemma:cond_coeff} are $v_1=-v_2$ and $w_1=w_2+v_1$. This defines 
the number field $\QQ(\zeta_9+\zeta_9^{-1})$, the maximal real subfield 
of the cyclotomic field $\QQ(\zeta_9)$.
Indeed, we have that $y \in 
\QQ(\zeta_9+\zeta_9^{-1})$ can be written as follows
\begin{eqnarray*}
y\!\!&\!\!\!=\!\!\!\!& y_0+y_1(\zeta_9+\zeta_9^{-1})+y_2(\zeta_9+\zeta_9^{-1})^2,
        ~y_i \in \QQ,~ i=0,1,2 \\
 &\!\!\! =\!\!\!\! & y_0 + y_1(\zeta_9-\zeta_9^2-\zeta_9^5)
            +y_2(2-\zeta_9+\zeta_9^2-\zeta_9^4) \\
 &\!\!\! =\!\!\!\! & (y_0+2y_2)+[(y_1-y_2)-y_2\zeta_3]\zeta_9 + 
                   [(y_2-y_1)-y_1\zeta_3]\zeta_9^2
\end{eqnarray*}
This finds ``formally'' a commutative subfield that we already 
found ``naturally''. 
\end{example}
Other examples can be found in Table \ref{table:poly}.
\begin{table}
\begin{center}
\begin{tabular}{|l|l|l|}
\hline
$\QQ(\zeta_3)(\nu)$ & minimal polynomial of $\nu$ & discriminant \\
\hline
$\nu=\theta+(1+\zeta_3)e-e^2$  & $X^3+X^2-5X-3$  & $2^2\cdot 3 \cdot 47$ \\
$\nu=2\theta+(1+\zeta_3)e-e^2$ & $X^3-X^2-12X+1$ & $11\cdot 659$ \\
$\nu=3\theta+(1+\zeta_3)e-e^2$ & $X^3-6X-1$      & $3^3\cdot 31$ \\
$\nu=4\theta+(1+\zeta_3)e-e^2$ & $X^3-11X+9$      & 3137 \\
$\nu=5\theta+(1+\zeta_3)e-e^2$ & $X^3-X^2-61X-13$ & $2^2\cdot 307 \cdot 727$ \\
\hline
\end{tabular}
\caption{Examples of commutative subfields of $\Ac$.}
\label{table:poly}
\end{center}
\end{table}

%
%****************************************************************************%
%
\subsection{Unitary matrices in $\Ac$}

We now illustrate how to build unitary matrices in the commutative subfield 
$\QQ(\zeta_9)/\QQ(\zeta_3)$ of $\Ac$ that we built in the previous subsection.
  
Take for example the element 
\begin{eqnarray*}
x &= & 1+\zeta_9 + \zeta_9^3 +\zeta_9^5 \in \QQ(\zeta_9)\\ 
  &= & (1+\zeta_3) + e + e^2\zeta_3 \in \Ac
\end{eqnarray*}
As a matrix, $x$ can be represented as 
\[
\xm=
\left(\begin{array}{ccc}
      1+\zeta_3 & \zeta_3^2  & \zeta_3 \\
      1         & 1+\zeta_3 & \zeta_3^2       \\
      \zeta_3   & 1         & 1+\zeta_3
      \end{array}
\right).
\]
We have
\begin{eqnarray*}
\alpha(x) &= & -\zeta_9^2 - \zeta_9^3 +\zeta_9^4 -\zeta_9^5 \in \QQ(\zeta_9)\\ 
          &= & -\zeta_3 + e\zeta_3 + e^2 \zeta_3^2 \in \Ac
\end{eqnarray*}
Again, as a matrix, $\alpha(x)$ can be represented as 
\[
\left(\begin{array}{ccc}
      -\zeta_3 &  1         & \zeta_3^2 \\
       \zeta_3 &  -\zeta_3  &  1       \\
      \zeta_3^2 & \zeta_3   & -\zeta_3
      \end{array}
\right)
\]
which can be checked to be $\xm^*$.
We have 
\[
x/\alpha(x)=1/19(-10+16\zeta_9+\zeta_9^2-4\zeta_9^3+14\zeta_9^4+8\zeta_9^5)
\] 
which has norm 1, so that by Corollary \ref{cor:equiv}, the matrix 
$\xm (\xm^*)^{-1}$ is unitary. This can be easily verified, since
\[
\begin{array}{c}
\xm (\xm^*)^{-1} = \\
\!\!\!\!\!\left(\begin{array}{ccc}
      -0.421-0.182i & 0.473+0.638i   & -0.157+0.36 i \\
      -0.236-0.319i &-0.421-0.182i   &  0.473+0.638i \\
      -0.789+0.09i  &-0.236-0.319i   & -0.421-0.182i
      \end{array}
\right)^T\!\!.
\end{array}
\]
Notice that this procedure can be applied
\begin{itemize}
\item 
to any element of $\QQ(\zeta_9)/\QQ(\zeta_3)$, and for each, 
it will give a unitary matrix.
There may obviously be some redundancy. Typically, if the element $x$ is 
invariant by $\alpha$, the above procedure will yield the identity matrix.
\item 
to any commutative subfield of $\Ac$, assuming that it has 
a quadratic subfield fixed by the involution, so that we obtain a family 
of unitary matrices for every such commutative subfields of the algebra. 
\item to any cyclic division algebra one starts with. 
\end{itemize}

\vspace{10cm}

%**************************************************************************%
%
% CONCLUSIONS
%
%**************************************************************************%

\section{Conclusion and future work}

In this paper, we showed how to build fully diverse matrices using 
cyclic division algebras. The main idea was to translate the condition of 
being unitary for a matrix into another condition that applies on  
commutative subfields of the algebra. We also showed that given a 
cyclic division algebra, several families of unitary matrices can be 
computed.
Unlike the property of being unitary, the full diversity naturally came 
by choosing $\Ac$ a division algebra. 

In this paper, we focussed on the existence of unitary matrices in division 
algebras. We did not investigate further the properties of these matrices.
In particular, we have not discussed yet their applications to noncoherent 
channel coding. Namely, we have not given explicit codebooks and analysis 
of their diversity. However, recent research work in that direction is 
already yielding promising results. In that perspective, there are several 
research topics we could suggest, among which: general lower bounds on the 
diversity, classification of the codebooks one can obtain from these 
families of unitary matrices, and in particular, criteria to design 
the ``best'' codebooks.

%******************************* Bibliography *****************************%

\end{document}